\documentclass[12pt]{article}

\usepackage{amsfonts,amssymb,amsmath}

\usepackage{graphicx}
\DeclareGraphicsExtensions{.pdf,.eps,.jpg}

\begin{document}
\title{\bf Enhancement of quantum transport efficiency in a noisy spin channel}
\author{ Naghi Behzadi $^{a}$ \thanks{E-mail:n.behzadi@tabrizu.ac.ir} ,
Bahram Ahansaz $^{b}$,
Abbas Ektesabi $^{a}$
\\ $^a${\small Research Institute for Fundamental Sciences, University of Tabriz, Iran,}
\\ $^b${\small Physics Department, Azarbaijan Shahid Madani University, Iran.}} \maketitle

\begin{abstract}
In this paper, based on the notion of pseudomode framework introduced by Garraway in [Phys. Rev. A 55, 2290 (1997)], we propose a mechanism for enhancing the efficiency of excitation energy transport in a spin channel which is in contact with a Lorentzian reservoir, through the inclusion of other similar auxiliary spin chains into the reservoir. To this aim, a Lindblad-type master equation for the dynamics of the transport process is provided on the basis of pseudomode approach. It is figured out that increasing the number of auxiliary chains in the reservoir enhances the efficiency of transport or equivalently, rises the population of the reaction center attached to the end of the channel. Moreover, it is interesting to note that the mechanism has better efficiency for the channels with longer lengths.
\end{abstract}

\section{Introduction}
Quantum transport of charge or energy plays a central role in many scientific disciplines and is the basis of many physical, chemical or biological processes.
The efficient transport of excitation energy through natural or artificial network of many-body interacting quantum systems has attracted the attentions of many researchers in recent years [1-7]. However, the dynamics of the quantum many-body systems is often investigated in the one-dimensional structures such as spin chains due to the existence of exact analytical solutions in some cases \cite{Takahashi} and efficient numerical methods for their simulation in other cases \cite{White}. Quantum spin chains are the best-known among the various physical systems that can serve as quantum channels and also can be realized in different physical systems such as arrays of Josephson junctions \cite{Tsomokos,Romito}, cold atoms in optical lattices \cite{Duan}, arrays of quantum dots \cite{Benjamin,Nikolopoulos} and etc.

In the context of quantum transport of excitation energy such as light harvesting (photosynthetic) complexes, manipulated dephasing noises have the ability to improve the efficiency of transport \cite{Behzadi1, zhang} by removing local or global destructive interferences of coherent hopping \cite{roben}. However, in general, any real quantum system suffers dissipation due to the unavoidable interaction with its surrendering environment \cite{Breuer, Gardiner} which ultimately causes the decay of excitation energy irreversibly into the environment. So, protecting the energy transport process from the noisy effects of the environment for improvement of its efficiency is a fundamental challenge in this field, as is in some other branches of quantum technology such as quantum information processing [20-24].

In this work, we propose a mechanism for enhancing the efficiency of excitation energy transport in a noisy spin channel which is in contact with a common Lorentzian dissipative reservoir. This procedure works through the inclusion of other similar auxiliary spin chains
into the reservoir when the system-environment coupling constants have been designed in a
particular way. In fact, by this designment, we can exploit the advantages of pseudomode
approach introduced in \cite{Garraway} to obtain an exact Lindblad-type master equation for the dynamics of the N-chain open system. Also,, since the addition of a sink site, as a reaction center, to the end of the spin channel makes it difficult to obtain an exact or perturbative master equation so, as illustrated later, we use the pseudomode approach to overcome this difficulty. We show that the inclusion of auxiliary chains into the reservoir enhances the efficiency of transport in the  channel and therefore, the population of the reaction center can be well-improved in this way. Also, it appears that the procedure has better performance for channels with longer lengths.

The paper is organized as follows: In Sec. II, the dynamics of $N$ identical spin chains interacting with a common structured reservoir is discussed by emphasizing that one of the chains is considered as our spin channel whose end site has been attached to an additional sink site. Sec. III is devoted to present the decisive role of the auxiliary chains on the enhancement of transport efficiency in the spin channel.
Finally, the paper is ended by a brief conclusion in Sec. IV.

\section{Dynamics of the system}
\subsection{The Hamiltonian}

In this section, we consider $N$ identical linear spin chains in a common structured reservoir, each of which consists of $M$ two-level system or qubit with uniform nearest-neighbor $XY$ interaction, as depicted in Fig. 1. The chains are non-interacting, i.e. they have no direct interaction with each other initially. The Hamiltonian of the whole system is considered as ($\hbar=1$)
\begin{eqnarray}
H=H_{S}+H_{R}+H_{SR},
\end{eqnarray}
where the system Hamiltonian $H_{S}$, the reservoir Hamiltonian $H_{R}$ and the system-reservoir interaction Hamiltonian $H_{SR}$
are respectively given by
\begin{eqnarray}
H_{S}=\omega_{0}\sum_{j=1}^{N} \sum_{l=1}^{M}\sigma_{l}^{j+}\sigma_{l}^{j-}+J\sum_{j=1}^{N}\sum_{l=1}^{M-1} \big( \sigma_{l}^{j+}\sigma_{l+1}^{j-}+\sigma_{l+1}^{j+}\sigma_{l}^{j-}\big),
\end{eqnarray}
\begin{eqnarray}
H_{R}=\sum_{k}\omega_{k}a_{k}^{\dag}a_{k},
\end{eqnarray}
and
\begin{eqnarray}
H_{SR}=\sum_{j=1}^{N}\sum_{l=1}^{M}\sum_{k} \big(g_{kl}^{j} \sigma_{l}^{j+}a_{k}+{g_{kl}^{j}}^{*}\sigma_{l}^{j-}a_{k}^{\dag} \big).
\end{eqnarray}
In these equations, $\omega_{0}$ is the transition frequency of two-level atoms and $J$ is the strength of uniform coupling between the nearest-neighbor atoms in the spin chains.
Also, $a_{k}$ ($a_{k}^\dag$) is the annihilation (creation) operator of the $k$th field mode with frequency $\omega_{k}$ and $g_{kl}^{j}$ is the coupling strength between the $k$th field mode and the $l$th atom of the $j$th chain. The lowering operator $\sigma_{l}^{j-}=(\sigma_{l}^{j+})^{\dagger}=|\mathbf{0}\rangle \langle (j-1)M+l|$ ($j=1,2,...,N$; $l=1,2,...,M$) describes decay from the excited state of $l$th atom of the $j$th chain into the ground state $|\mathbf{0}\rangle=|0\rangle^{\otimes MN}$, where
\begin{eqnarray}
\big|(j-1)M+l\big\rangle=|0\rangle^{\otimes M}...|0\rangle^{\otimes M}\otimes{\underbrace{|\overbrace{0...0}^{l-1} 1 \overbrace{0...0}^{M-l}\rangle}_{jth}}\otimes|0\rangle^{\otimes M}...|0\rangle^{\otimes M},
\end{eqnarray}
is a typical standard basis state for the $MN$-dimensional single excitation subspace of the chains.
In the following, we define another set of basis in terms of the standard basis (5), as
\begin{eqnarray}
|\varphi_{j}^{l}\rangle=\sqrt{\frac{2}{M+1}} \sum_{i=1}^{M} \mathrm{sin}(iq_{l}) \big|(j-1)M+i\big\rangle=\sum_{i=1}^{M} u_{li} \big|(j-1)M+i\big\rangle, \quad q_{l}=\frac{\pi l}{M+1},
\end{eqnarray}
which are eigenstates of the $H_{S}$ with corresponding eigenvalues $E_{l}=\omega_{0}+2J\mathrm{\cos}(q_{l})$ and $u_{li}=\sqrt{\frac{2}{M+1}}\mathrm{sin}(iq_{l})$. It should be noted that the eigenstates (6) are reduced to the same ones of Ref. $\cite{Christandl1}$ for a single chain ($N=1$). Now, we design the system-reservoir coupling in a particular way such that $g_{kl}^{j}=G_{k}^{j} u_{rl}$ with $r=1,2,...,M$, so the total Hamiltonian (1) in the new basis (6) takes the following form

\begin{eqnarray}
H=\sum_{j=1}^{N}\sum_{l=1}^{M}E_{l}|\varphi_{j}^{l}\rangle \langle\varphi_{j}^{l}|+
\sum_{k}\omega_{k} a_{k}^{\dag}a_{k}+\sum_{j=1}^{N}\sum_{k} \big( G_{k}^{j} |\varphi_{j}^{r}\rangle \langle\mathbf{0}| a_{k}+{G_{k}^{j}}^{*} |\mathbf{0}\rangle \langle\varphi_{j}^{r}| a_{k}^{\dag}\big).
\end{eqnarray}
Under these considerations, we find that in Eq. (7), interaction of the system with the reservoir takes place only through the $N$ eigenstates $\{|\varphi_{1}^{r}\rangle, |\varphi_{2}^{r}\rangle, ..., |\varphi_{N}^{r}\rangle \}$ (for a particular value of $r$). In fact, these eigenstates are the degenerated excited states of a $(N+1)$-level $V$-type configuration with ground state $|\mathbf{0}\rangle$ and transition frequency $E_{r}=\omega_{0}+2J\mathrm{\cos}(q_{r})$ corresponding to all transitions $|\varphi_{j}^{r}\rangle \leftrightarrow |\mathbf{0}\rangle$ ($j=1,2,..,N$), as shown in Fig. 2. Therefore, the other $MN-N$ eigenstates $\{|\varphi_{1}^{l}\rangle, |\varphi_{2}^{l}\rangle, ..., |\varphi_{N}^{l}\rangle \}$ with $l\neq r$ are decoupled from the reservoir.

Now let us assume that one of the chains is considered as our spin channel under study, and the remaining ones are used as auxiliary chains. By considering the Schrodinger equation for the Hamiltonian (7), and that the reservoir spectral density is Lorentzian, the exact dynamics for the $N$-chain open system (as well as for the spin channel) can be obtained in a similar way of Ref. \cite{beh} carried out for a $3$-level $V$-type atom. However, in general, addition of a sink site as a reaction center to the end of the spin channel causes difficulties in obtaining the useful dynamics in order to describe the excitation transport from the initial site of the channel into the reaction center. In other words, it is not only impossible to obtain an exact master equation for the process of excitation transport into the reaction center, but also exploiting an alternative perturbative Lindblad-type master equation is so difficult. Nevertheless, from the view point of the pseudomode approach \cite{Garraway}, there exist an exact Lindblad-type master equation for the dynamics of a multilevel $V$-type atomic system interacting with a Lorentzian structured reservoir. In fact, the pseudomode approach enables us to include another Lindblad term corresponding to the irreversible dissipation process arisen from the presence of a reaction center attached to the end of the channel. Therefore, an efficient framework can be provided for analysing the excitation transport process, along with the discussion about the improvement of its efficiency, on the basis of pseudomode approach which will be illustrated in the next subsection.

\subsection{The pseudomode approach}

For a given reservoir, the pseudomodes are auxiliary variables being introduced in one-to-one correspondence
with the poles of the respective spectral distribution function in the complex frequency plane. This approach allows us to derive an exact master equation without using the perturbation theory, Born or Markov approximation. Such an exact master equation
describes the coherent interaction between the system and the pseudomodes in the presence of
decay of the pseudomodes due to the interaction with a Markovian reservoir. For example,
this approach has been utilized to develop an equivalent master equation for a three-level $V$-type system \cite{Garraway1}, multilevel $V$-type \cite{Garraway}, and
a three-level atom in a ladder configuration \cite{Dalton}. Also, the exact entanglement dynamics of a two-qubit system  which is in contact with a common structured reservoir has been studied by using the pseudomode mechanism through  establishing a connection with a three-level ladder system \cite{Mazzola}.

In the following, we briefly review the key features of the pseudomode theory which has been utilized to derive an exact Lindblad-type master equation for the dynamics of a multilevel $V$-type atomic system, as discussed in \cite{Garraway}. To this aim, we consider a multilevel $V$-type atomic system coupled to a bath of harmonic oscillators.
This atom is consisting of a single ground-state $|0\rangle_{a}$ coupled to the number of upper excited states such as $|i\rangle_{a}$ ($i=1,2,..., N$) with the corresponding energy difference $\omega_{i}$.
Thus, the Hamiltonian of the system is given by ($\hbar=1$)
\begin{eqnarray}
H'=\sum_{i=1}^{N}\omega_{i}|i\rangle_{a a}\langle i|+\sum_{\lambda}\omega_{\lambda}a_{\lambda}^{\dagger}a_{\lambda}+\sum_{i=1}^{N}\sum_{\lambda} \big(  \mathfrak{g}_{i\lambda} |i\rangle_{aa} \langle 0| a_{\lambda}+\mathfrak{g}_{i\lambda}^{*}|0\rangle_{aa} \langle i|a_{\lambda}^{\dagger}\big),
\end{eqnarray}
where $a_{k}$ and $a_{k}^\dag$ are the respective annihilation and creation operators for each oscillation mode of the reservoir with frequency $\omega_{\lambda}$, and $\mathfrak{g}_{i\lambda}$ is the coupling strength of the $i$th excited state of the multilevel atom and the $\lambda$th mode of the reservoir.
To describe a one photon excitation process, we work in the single excitation subspace where the state vector of the total system can be written as
\begin{eqnarray}
|\psi(t)\rangle=c_{0}(t) |0\rangle_{a}|0\rangle_{e}+ \sum_{i}c_{i}(t)|i\rangle_{a}|0\rangle_{e}+\sum_{\lambda}c_{\lambda}(t)|0\rangle_{a}|1_{\lambda}\rangle_{e}.
\end{eqnarray}
Here, the ket $|0\rangle_{e}$ shows that all of the reservoir modes are in their respective vacuum states while the
ket $|1_{\lambda}\rangle_{e}$ indicates that all of the reservoir modes, except the $\lambda$ one, are in the ground states.

Substitution Eq. (8) and Eq. (9) into the time dependent Schrodinger equation $id/dt|\psi(t)\rangle=H'|\psi(t)\rangle$, leads to the following infinite set of coupled differential equations
\begin{equation}
\begin{array}{l}
\displaystyle i\frac{dc_{i}(t)}{dt}=\omega_{i}c_{i}(t)+\sum_{\lambda}\mathfrak{g}_{i\lambda}c_{\lambda}(t),\\\\
\displaystyle i\frac{dc_{\lambda}(t)}{dt}=\omega_{\lambda}c_{\lambda}(t)+\sum_{i}\mathfrak{g}_{i\lambda}^{*}c_{i}(t).
\end{array}
\end{equation}
It is clear that the coefficient $c_{0}(t)$ will be invariant in time, which means $c_{0}(t)=c_{0}(0)$.
Formally, eliminating the coefficient $c_{\lambda}(t)$ enables us to derive an integro-differential equation for the amplitudes of the excited states of the atom, i.e. $c_{i}(t)$s. After some algebraic manipulations, a closed set of integro-differential equations for $c_{i}(t)$, $i=1, 2, 3, ..., N$, is obtained as
\begin{eqnarray}
\frac{dc_{i}(t)}{dt}=-i\omega_{i}c_{i}(t)-\int_{0}^{t} dt'\sum_{j} G_{ij}(t-t') c_{j}(t'),
\end{eqnarray}
where the kernel function $G_{ij}(t-t')$ can be expressed in terms of the spectral density $J_{ij}(\omega)$ of the reservoir as follows
\begin{eqnarray}
G_{ij}(t-t')=\int_{-\infty}^{\infty}d\omega J_{ij}(\omega) e^{-i\omega(t-t')},
\end{eqnarray}
with
\begin{eqnarray}
J_{ij}(\omega)=\frac{\Omega_{i} \Omega_{j}}{2\pi} D(\omega).
\end{eqnarray}
Here, $\Omega_{i}$ ($\Omega_{i}^2= \sum_{\lambda} |\mathfrak{g}_{i\lambda}|^2$) is the coupling strength between the $i$th excited state of the $(N+1)$-level atom and the reservoir,
and $D(\omega)$ is the reservoir structure function with the normalization $\int_{-\infty}^{\infty}d\omega D(\omega)=2\pi$, which enables us to consider the various types of the reservoir.

It is very useful to calculate the integral (12) from a contour in the complex $\omega$ plane which is closed in the lower half plane due to the vanishing of the exponential part of the integrand (because $t>t'$). So, taking a contour in the lower half plane gives
\begin{eqnarray}
G_{ij}(t-t')=-\frac{\Omega_{i} \Omega_{j}}{2\pi} \oint_{C} dz D(z) e^{-iz(t-t')}.
\end{eqnarray}
We suppose that the function $D(z)$ has poles in the lower half plane at $z_{1},z_{2},...,z_{l},...$ and the residues of $D(z)$ are denoted by $r_{1},r_{2},...,r_{l},...$.
Then, by using the Theorem of Residues, we have
\begin{eqnarray}
G_{ij}(t-t')=-i \Omega_{i} \Omega_{j} \sum_{l} r_{l} e^{-iz_{l}(t-t')}.
\end{eqnarray}
As a result, the integral in Eq. (12) can be converted into a sum over $l$ which makes it easier to calculate if we know the positions and residues
of the poles of $D(z)$. Each of the poles of the reservoir in the lower half complex $\omega$ plane will be associated with one pseudomode.
In the following, by inserting Eq. (15) into the Eq. (11), we find that
\begin{eqnarray}
i\frac{dc_{i}(t)}{dt}=\omega_{i}c_{i}(t)-\sum_{l} \Omega_{i}r_{l} \sum_{j} \Omega_{j}e^{-iz_{l}t} \int_{0}^{t} dt'e^{iz_{l}t'} c_{j}(t').
\end{eqnarray}
Now, based on Eq. (16), we can introduce a fictional pseudomode amplitude as
\begin{eqnarray}
s_{l}(t)=-i\sum_{j} \Omega_{j} \sqrt{-ir_{l}} e^{-iz_{l}t} \int_{0}^{t} dt'e^{iz_{l}t'} c_{j}(t').
\end{eqnarray}
So, Eq. (11) can be converted into the following form
\begin{equation}
\begin{array}{l}
\displaystyle i\frac{dc_{i}(t)}{dt}=\omega_{i}c_{i}(t)+\sum_{l}\mathfrak{g}_{il}s_{l}(t),\\\\
\displaystyle i\frac{ds_{l}(t)}{dt}=z_{l}s_{l}(t)+\sum_{i}\mathfrak{g}_{il}c_{i}(t),
\end{array}
\end{equation}
where the last equation follows from the differentiation of Eq. (17) and the coupling between the pseudomode $l$ and the atomic level $i$ is $\mathfrak{g}_{il}=\Omega_{i} \sqrt{-ir_{l}}$.
Furthermore, it should be noted that Eq. (12) implies that $G_{ij}(0)=\Omega_{i}\Omega_{j}$ and thus by considering Eq. (15), we always have
\begin{eqnarray}
\sum_{l} (-ir_{l})=1.
\end{eqnarray}
Therefore, the original problem consisted of an infinite set of ordinary differential equations (10) have been converted into
a finite set (18). Consequently, the Eqs. (18) for the enlarged system consisting of the atom plus the pseudomode can now take the Lindblad-type master equation \cite{Garraway}
\begin{eqnarray}
\frac{\partial \rho}{\partial t}=-i[H_{0},\rho]+\sum_{l} (L_{l} \rho L_{l}^{\dag}-\frac{1}{2}L_{l}^{\dag}L_{l} \rho -\frac{1}{2}\rho L_{l}^{\dag}L_{l}),
\end{eqnarray}
where $\rho$ is the density matrix of the system comprising the pseudomodes and the atom and $H_{0}$ is the Hermitian Hamiltonian defined as
\begin{eqnarray}
H_{0}=\sum_{i=1}^{N}\omega_{i}|i\rangle_{a a}\langle i|+\sum_{l} Re(z_{l}) b_{l}^{\dag}b_{l}+\sum_{i=1}^{N}\sum_{l} \mathfrak{g}_{il} \big(|i\rangle_{aa} \langle 0|b_{l}+
|0\rangle_{aa} \langle i|b_{l}^{\dagger}\big).
\end{eqnarray}
In Eq. (21) the operator $b_{l}$ ($b_{l}^{\dag}$) is the annihilation (creation) operator for the excitation of the fictional mode $l$.
Also, the Lindblad operators involving the pseudomodes take the form
\begin{eqnarray}
L_{l}=\sqrt{-2\mathrm{Im}(z_{l})}b_{l}.
\end{eqnarray}

In the following, we consider the most straightforward example of a single pseudomode given by the Lorentzian spectral distribution with the structure function
\begin{eqnarray}
D(\omega)=\frac{\Gamma}{(\omega-\omega_{c})^2+(\Gamma/2)^2},
\end{eqnarray}
which has a single pole in the lower half complex frequency plane as
\begin{eqnarray}
z_{1}\equiv \omega_{c}-i\frac{\Gamma}{2},
\end{eqnarray}
where the constants $\Gamma$ and $\omega_{c}$ are the decay rate and frequency of the pseudomode respectively. Since we have a single pole at $z_{1}$,
so the normalization property in Eq. (19) gives $(-ir_{1})=1$. As a result, the single pseudomode coupling is a real quantity and it is given by $\mathfrak{g}_{i1}=\Omega_{i}\sqrt{-ir_{1}}=\Omega_{i}$.
Now, exploiting the pseudomode approach for our system of $N$ chains contained in a Lorentzian structured reservoir gives the following master equation
\begin{eqnarray}
\frac{\partial \widetilde{\rho}}{\partial t}=-i[\mathbb{H},\widetilde{\rho}]+\frac{\Gamma}{2}(2b \widetilde{\rho} b^{\dag}-b^{\dag}b \widetilde{\rho} -\widetilde{\rho} b^{\dag}b) \equiv \mathfrak{L}(\widetilde{\rho}),
\end{eqnarray}
where $\widetilde{\rho}$ is the density matrix of the extended system, i.e., the system comprising the chains and the pseudomode, and $\mathbb{H}$ is the Hermitian Hamiltonian
given in Eq. (21) as
\begin{eqnarray}
\mathbb{H}=\sum_{j=1}^{N}\sum_{l=1}^{M} E_{l}|\varphi_{j}^{l}\rangle \langle\varphi_{j}^{l}|+
\omega_{c} b^{\dag}b+\sum_{j=1}^{N} \Omega_{j} \big(|\varphi_{j}^{r}\rangle \langle\mathbf{0}| b+|\mathbf{0}\rangle \langle\varphi_{j}^{r}| b^{\dag}\big).
\end{eqnarray}
Here, $\Omega_{j}$ is corresponding to the coupling strength between the single pseudomode and the $j$th excited state of the $(N+1)$-level $V$-type system.

\section{Results}

In the previous section, we obtained a Lindblad-type master equation for the dynamics of the $N$ identical chains interacting with a common Lorentzian reservoir, in an analogous way of interaction of the $(N+1)$-level $V$-type system. In this section, based on the pseudomode framework, we explicitly investigate improving the efficiency of excitation transport in the spin channel, specified namely by $j=1$, through the inclusion of $N-1$ auxiliary chains into the reservoir. The first site of the channel has been initially excited, while whose $M$th site has been attached to the sink site or reaction center (see Fig. 1). The Lindblad-type master equation for the transport of excitation in the spin channel is as follows
\begin{eqnarray}
\frac{\partial \widetilde{\rho}}{\partial t}=\mathfrak{L}(\widetilde{\rho})+\mathfrak{L}_{\mathrm{sink}}(\widetilde{\rho}),
\end{eqnarray}
where $\widetilde{\rho}$ is the density matrix of the extended system comprises the chains, sink site and the pseudomode. The term $\mathfrak{L}(\widetilde{\rho})$ is the same as the Eq. (25), while $\mathfrak{L}_{\mathrm{sink}}(\widetilde{\rho})$ describes the population of the sink site
through an irreversible decay process from the chosen $M$th site of the channel, which is given by
\begin{eqnarray}
\mathfrak{L}_{\mathrm{sink}}(\widetilde{\rho})=\frac{\Gamma_{\mathrm{sink}}}{2}\Big(2\sigma^{+}_{\mathrm{sink}} \sigma^{1-}_{M} \widetilde{\rho} \sigma^{1+}_{M} \sigma^{-}_{\mathrm{sink}}-
\{\sigma^{1+}_{M} \sigma^{-}_{\mathrm{sink}} \sigma^{+}_{\mathrm{sink}} \sigma^{1-}_{M} ,\widetilde{\rho}\}\Big).
\end{eqnarray}
In Eq. (28), $\Gamma_{\mathrm{sink}}$ is the rate of the dissipative process that reduces the number of excitations in the
channel and traps it in the sink, and $\sigma^{+}_{\mathrm{sink}}$ and $\sigma^{-}_{\mathrm{sink}}$ are the raising and lowering operators of the sink site, respectively. By integrating from Eq. (27), the population of the sink or efficiency of transport is obtained as
\begin{eqnarray}
P_{\mathrm{sink}}(t)=2\Gamma_{\mathrm{sink}}\int_{0}^{t}\rho_{M, M}(t')dt'.
\end{eqnarray}

Now, our main claim is that the efficiency of transport of excitation from the initial site of the channel to the reaction center can be improved by entering the similar auxiliary chains into the reservoir. In other words, the efficiency of transport in the spin channel can be well-controlled using the auxiliary chains.

Numerical analysis demonstrates the performance of the introduced protocol. Figs. 3 and 4, show the population of the sink for the channels with $M=3,5$ qubits. It is observed, for the channel with $M=3$ qubits, that the quality of the quantum transport is seriously degraded due to the dissipation of the excitation energy throughout the channel in the absence of the auxiliary chains ($N=1$), as shown in Fig. 3. While by entering the auxiliary three-qubit chains ($N=2,6$) into the reservoir the efficiency of transport is considerably improved (Fig. 3). Similar behaviors are taken place for the case of the five-qubit channel (see Fig. 4). It is interesting to note that the protocol has better efficiency for the channels with longer lengths. This can be easily illustrated using the fact that, in the Hamiltonian (7) and (26), the interaction of the $N$ chains with the reservoir is established only through the $N$-dimensional subspace (the excited states of the ($N+1$)-level $V$-type configuration), so we have a noise-free subspace of dimension $(M-1)N$ which increases by increasing the length of the chains.

\section{Conclusions}
We presented a method for enhancing the efficiency
of transport of excitation energy in a noisy spin channel. It was indicated that the transport process in such a  channel, whose end has been attached irreversibly to a reaction center and is in contact with a dissipative Lorentzian structured reservoir, can be well-described by a Lindblad-type master equation obtained using the pseudomode approach.
By using this framework, we observed as a result that the efficiency of transport is well-improved through the addition of some other similar auxiliary spin chains into the reservoir. Moreover, we showed that this procedure has better efficiency for the channels with longer lengths.

\newpage
Fig. 1. A schematic representation of a spin channel in the presence of, e.g., four
similar auxiliary chains each of which has $M=4$ spins, and all of them are contained in a common structured reservoir (the yellow circle).
The first site of the channel is initially populated (the orange site), and the green arrow indicates an irreversible transfer of excitation from the last site of the channel into the sink.

\begin{figure}
\centering
\includegraphics[width=250 pt]{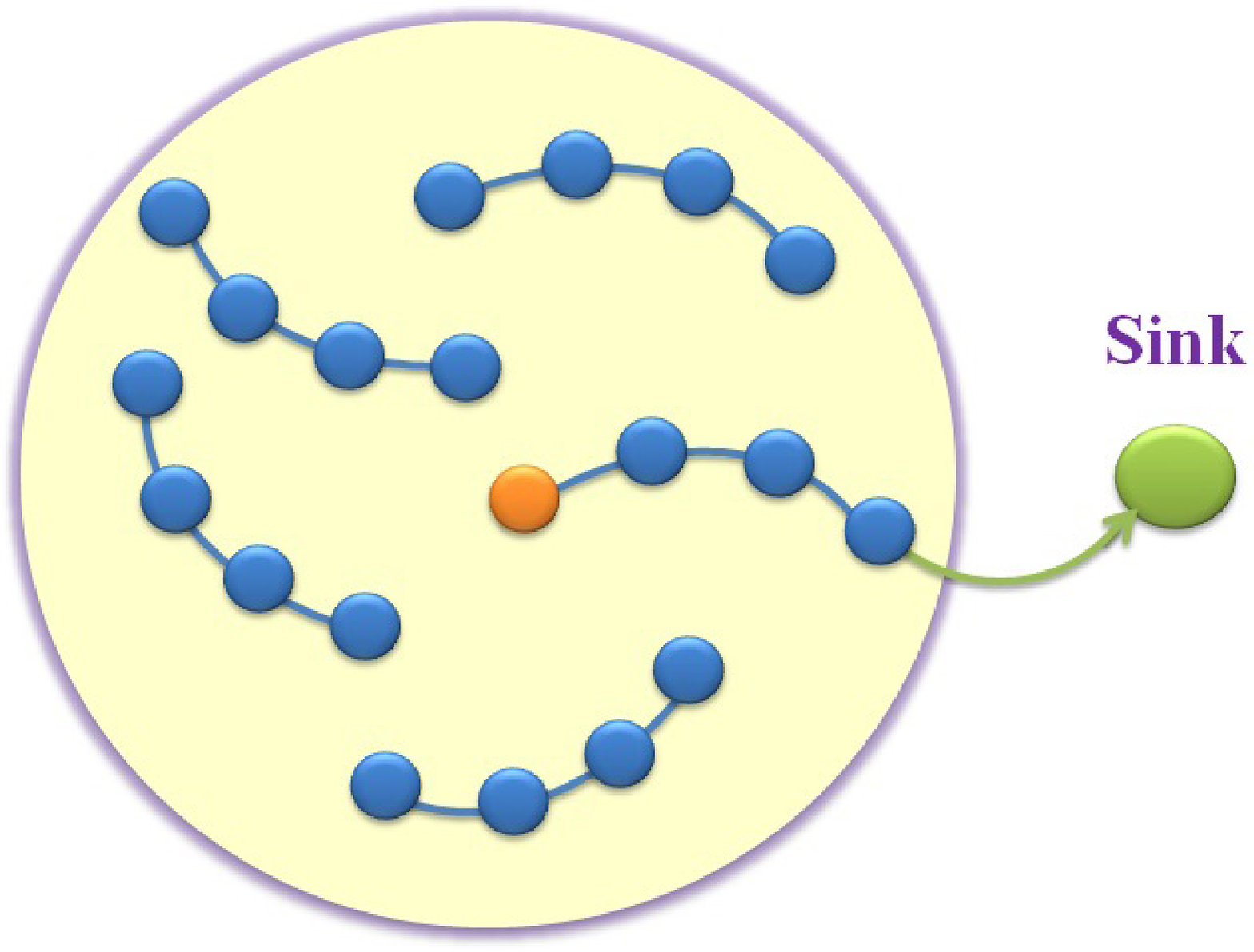}
\caption{} \label{fig2}
\end{figure}

\newpage
Fig. 2. The $(N+1)$-level $V$-type atomic system with ground state $|\mathbf{0}\rangle$ and $N$ degenerated excited states $\{|\varphi_{1}^{1}\rangle, ..., |\varphi_{N}^{1}\rangle \}$ with the transition frequency $E_{1}$ (in Eq. (7), $r$ is taken to be 1).

\begin{figure}
\centering
\includegraphics[width=300 pt]{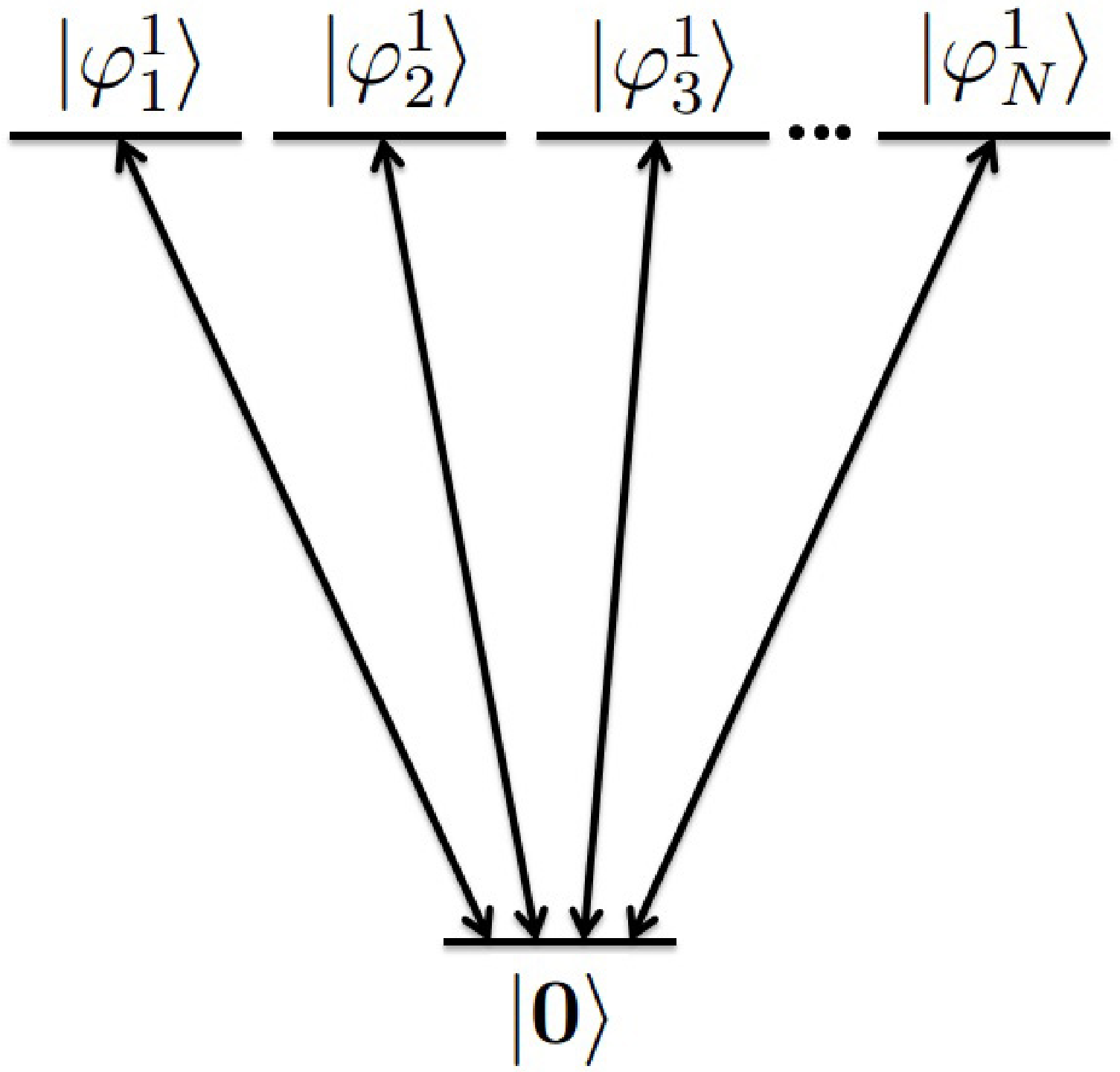}
\caption{} \label{fig2}
\end{figure}

\newpage
Fig. 3. Population of the sink (or efficiency of transport) in terms of time (in units of $\omega_{0}^{-1}$) for a spin channel of length $M=3$ spins, without use of the auxiliary chains, i.e., $N=1$ (the solid green curve), using of one auxiliary chain, i.e. $N=2$ (the dotted-dashed blue curve) and using of five auxiliary chains, i.e. $N=6$ (the dashed red curve). These panels are plotted with the fixed parameters $\omega_{c}=1.02$ (in units of $\omega_{0}$), $\Omega_{0}=0.15$ (in units of $\omega_{0}$) and $\Gamma_{sink}=0.6$ (in units of $\omega_{0}$).

\begin{figure}
        \qquad \qquad\qquad\qquad \qquad a \qquad\qquad \quad\qquad\quad\quad\qquad\qquad\qquad\qquad\qquad b\\{
        \includegraphics[width=3.3in]{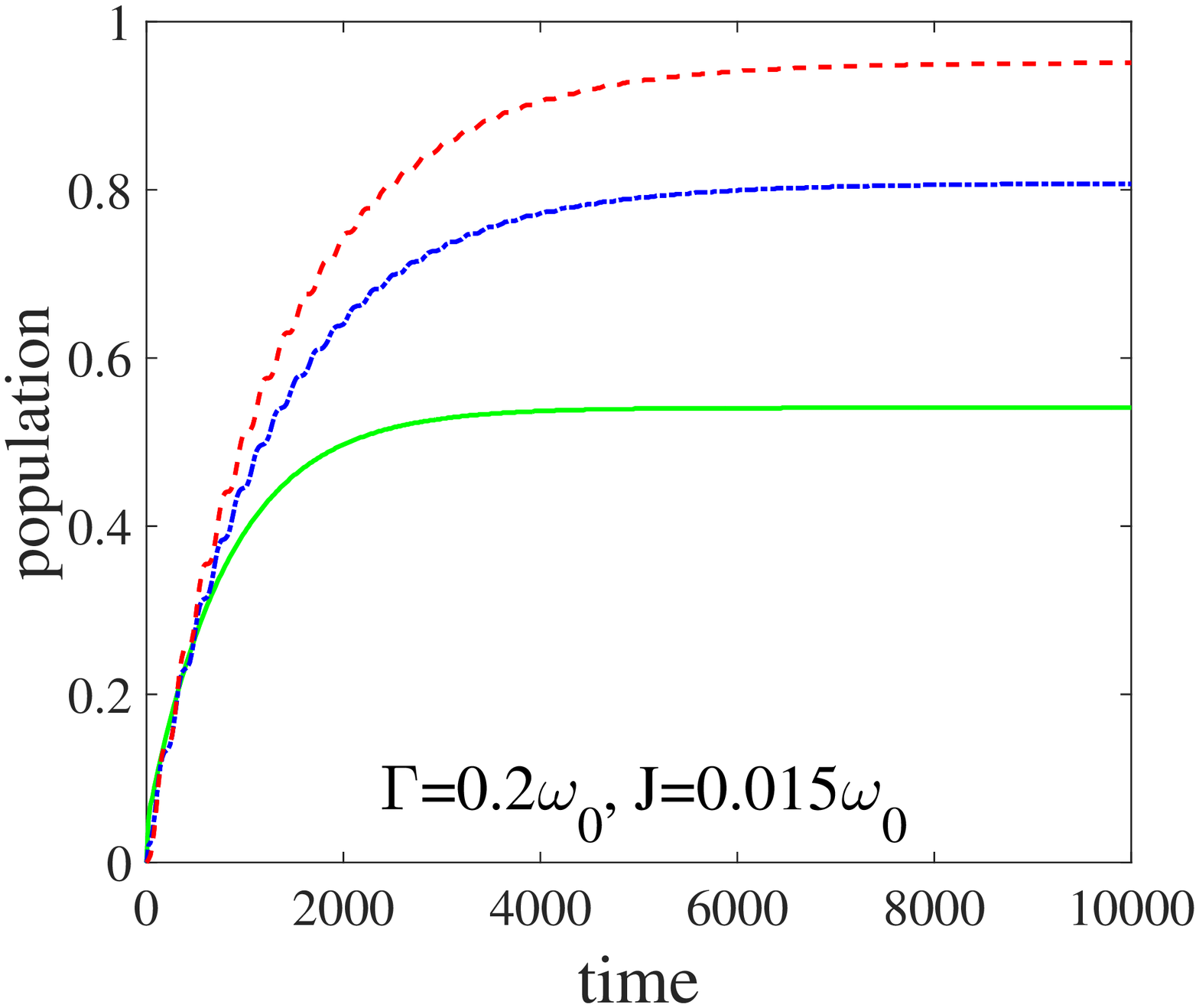}
        \label{fig:first_sub}
    }{
        \includegraphics[width=3.3in]{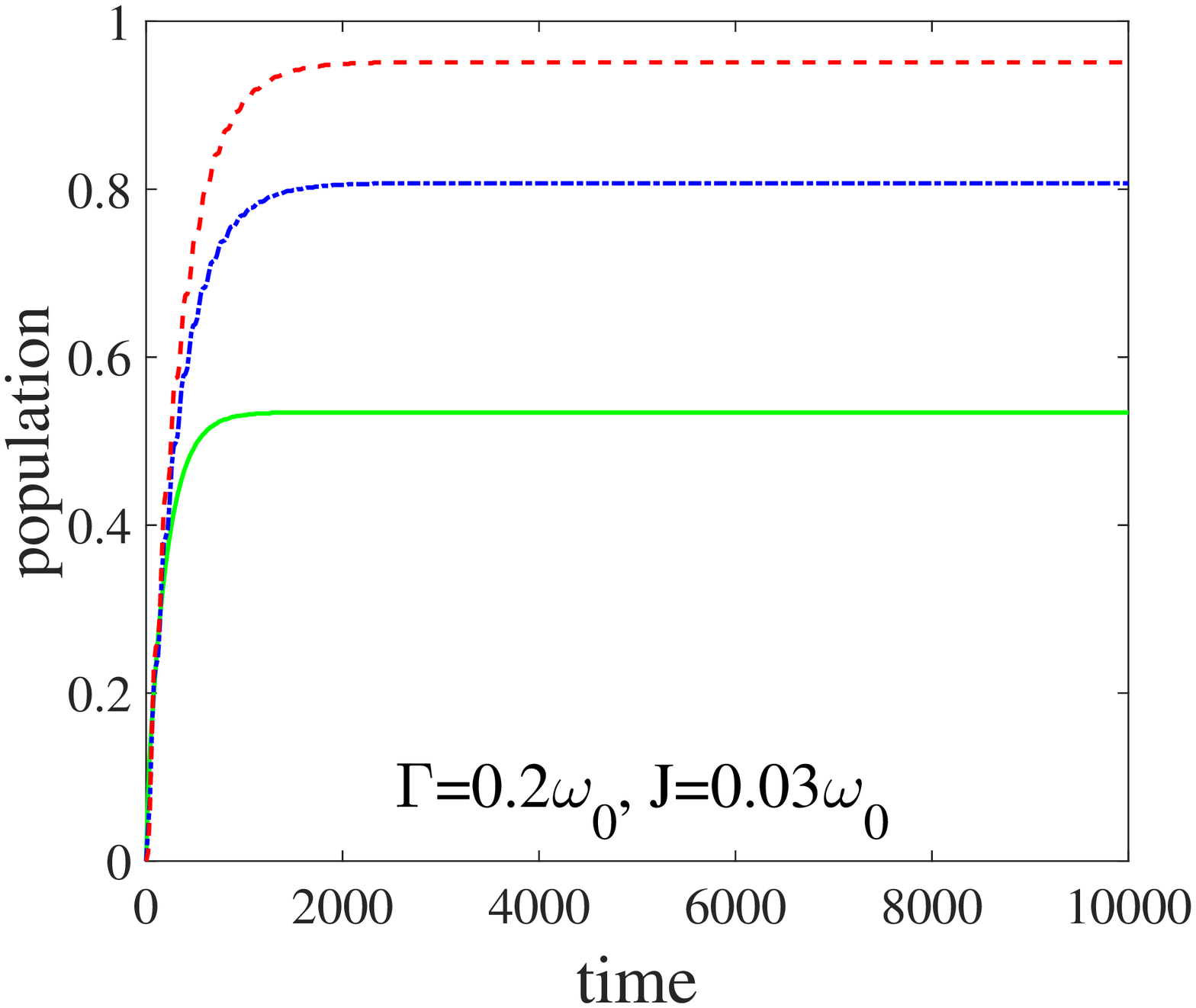}
        \label{fig:second_sub}
    }\\ \par \quad \quad\qquad\qquad\qquad \qquad c \qquad \qquad\quad\quad\qquad\qquad\qquad\quad\qquad\qquad\qquad d\\{
        \includegraphics[width=3.3in]{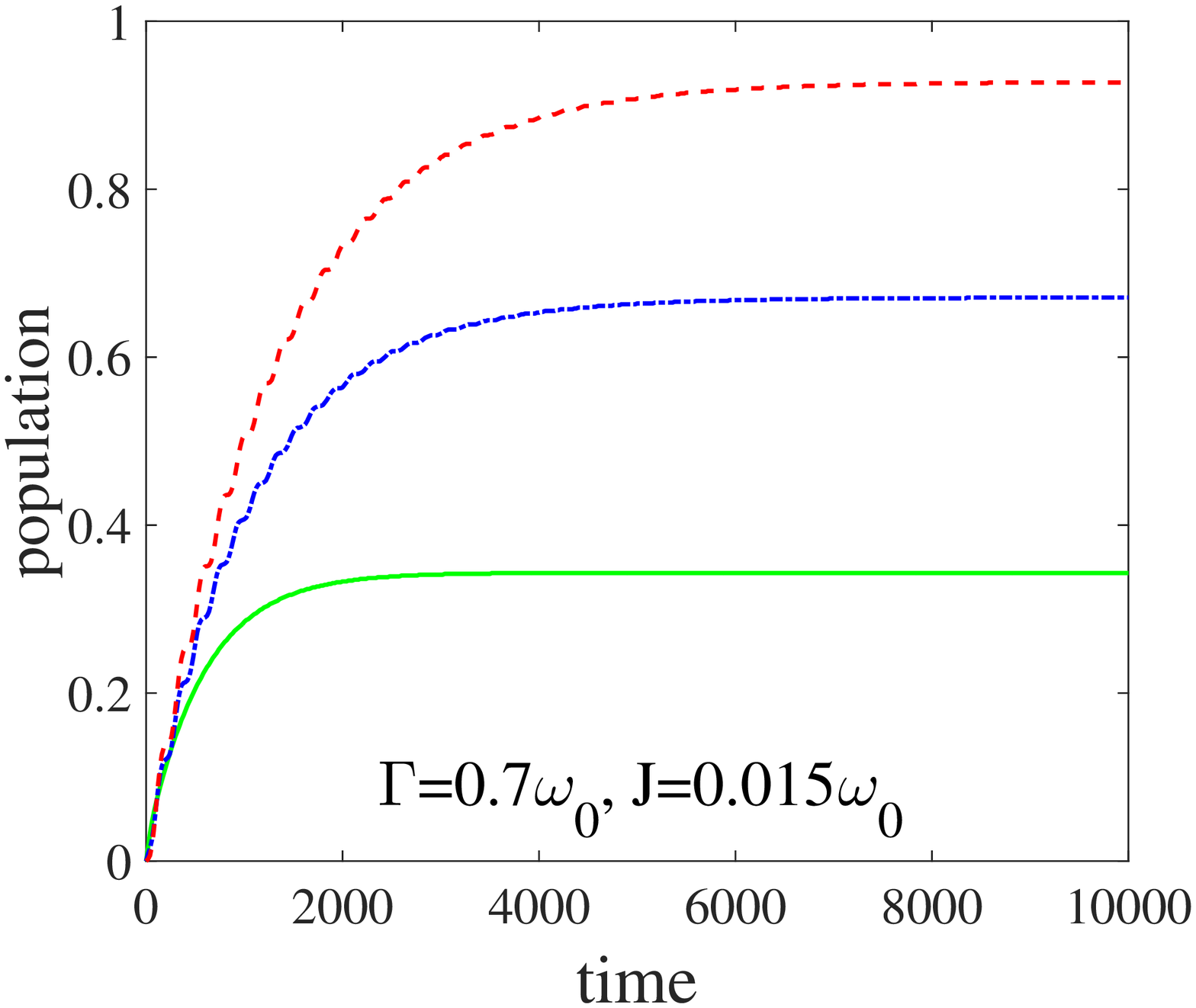}
        \label{fig:first_sub}
    }{
        \includegraphics[width=3.3in]{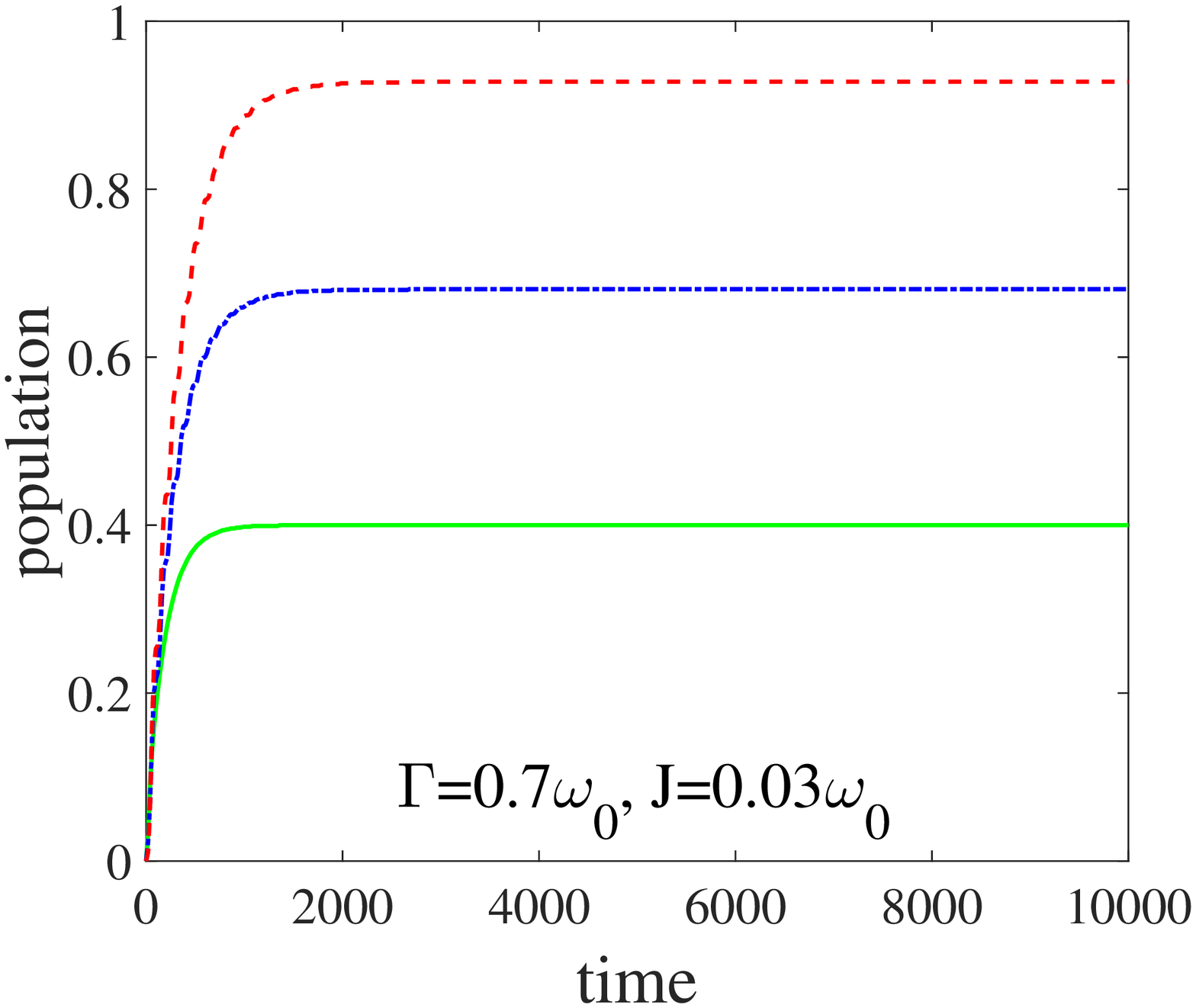}
        \label{fig:second_sub}
    }
    \caption{}
    \end{figure}

\newpage
Fig. 4. Population of the sink (or efficiency of transport) in terms of time (in units of $\omega_{0}^{-1}$) for a spin channel of length $M=5$ spins, without use of the auxiliary chains, i.e. $N=1$ (the solid green curve), using of one auxiliary chain, i.e. $N=2$ (the dotted-dashed blue curve) and using of five auxiliary chains, i.e., $N=6$ (the dashed red curve). These panels are plotted with the fixed parameters $\omega_{c}=1.02$ (in units of $\omega_{0}$), $\Omega_{0}=0.15$ (in units of $\omega_{0}$) and $\Gamma_{sink}=0.6$ (in units of $\omega_{0}$).

\begin{figure}
        \qquad \qquad\qquad\qquad \qquad a \qquad\qquad \quad\qquad\qquad\qquad\qquad\qquad\qquad\quad\quad b\\{
        \includegraphics[width=3.3in]{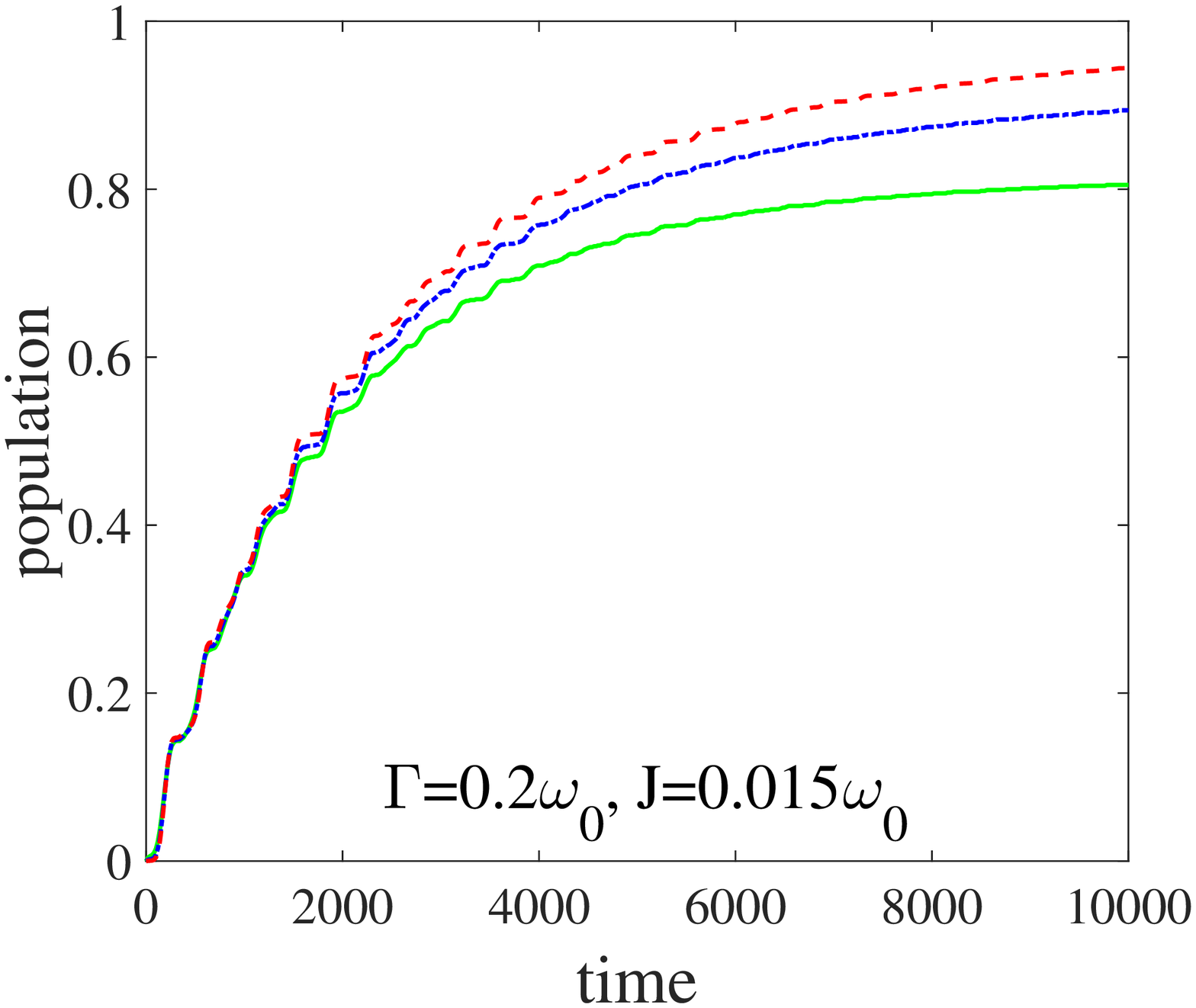}
        \label{fig:first_sub}
    }{
        \includegraphics[width=3.3in]{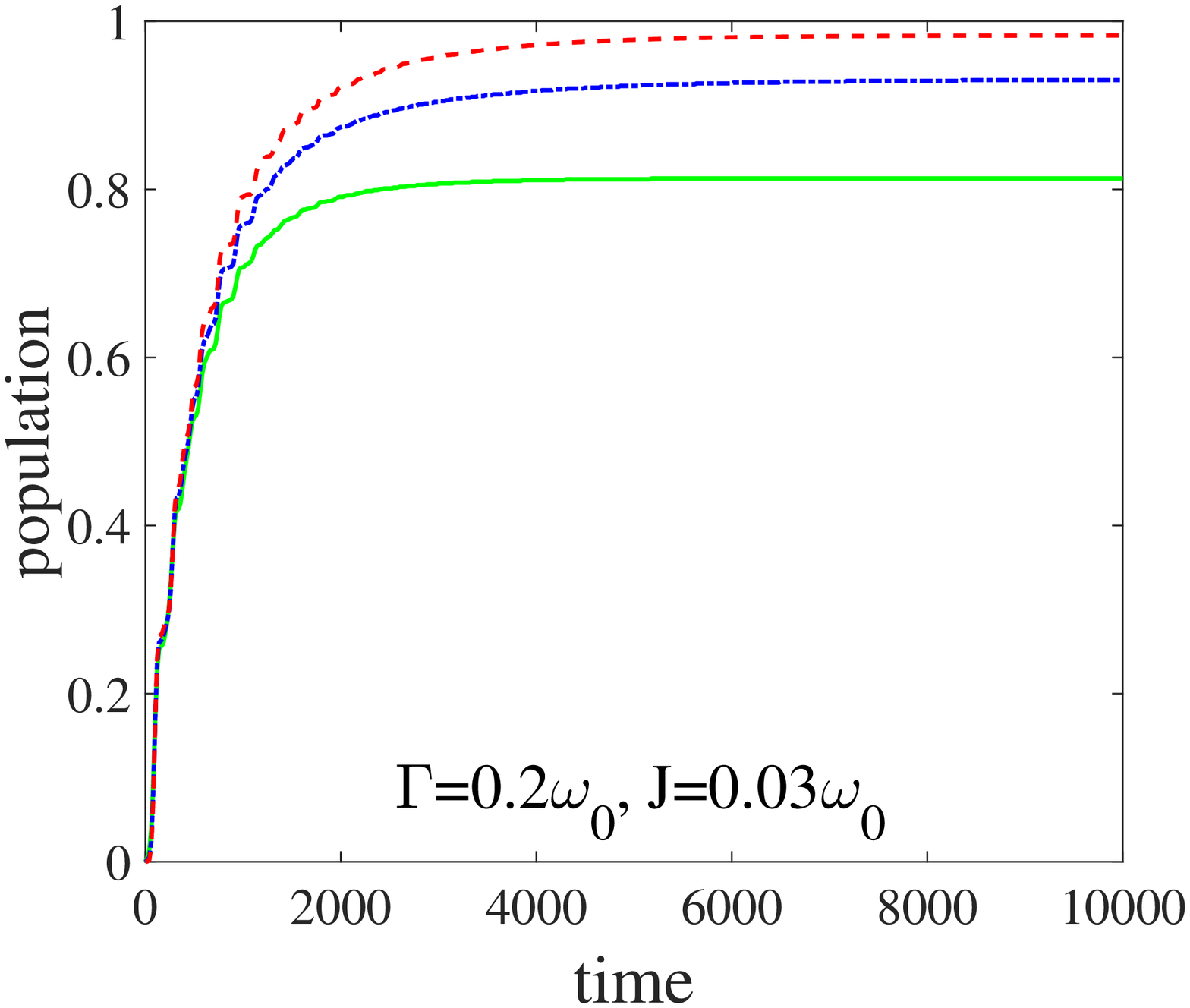}
        \label{fig:second_sub}
    }\\ \par \quad \quad\qquad\qquad\qquad \qquad c \qquad \qquad\qquad\quad\quad\qquad\qquad\quad\qquad\qquad\qquad d\\{
        \includegraphics[width=3.3in]{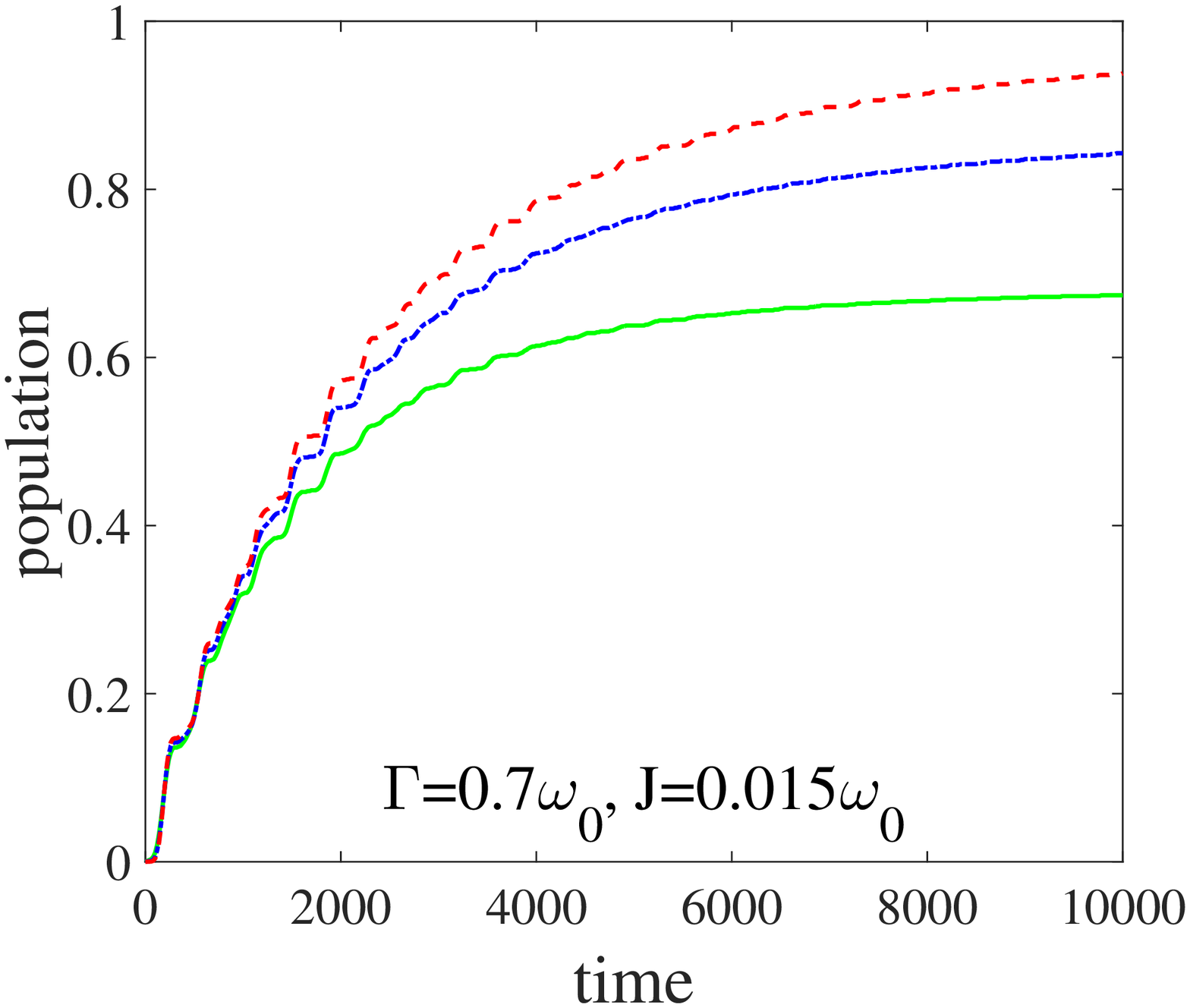}
        \label{fig:first_sub}
    }{
        \includegraphics[width=3.3in]{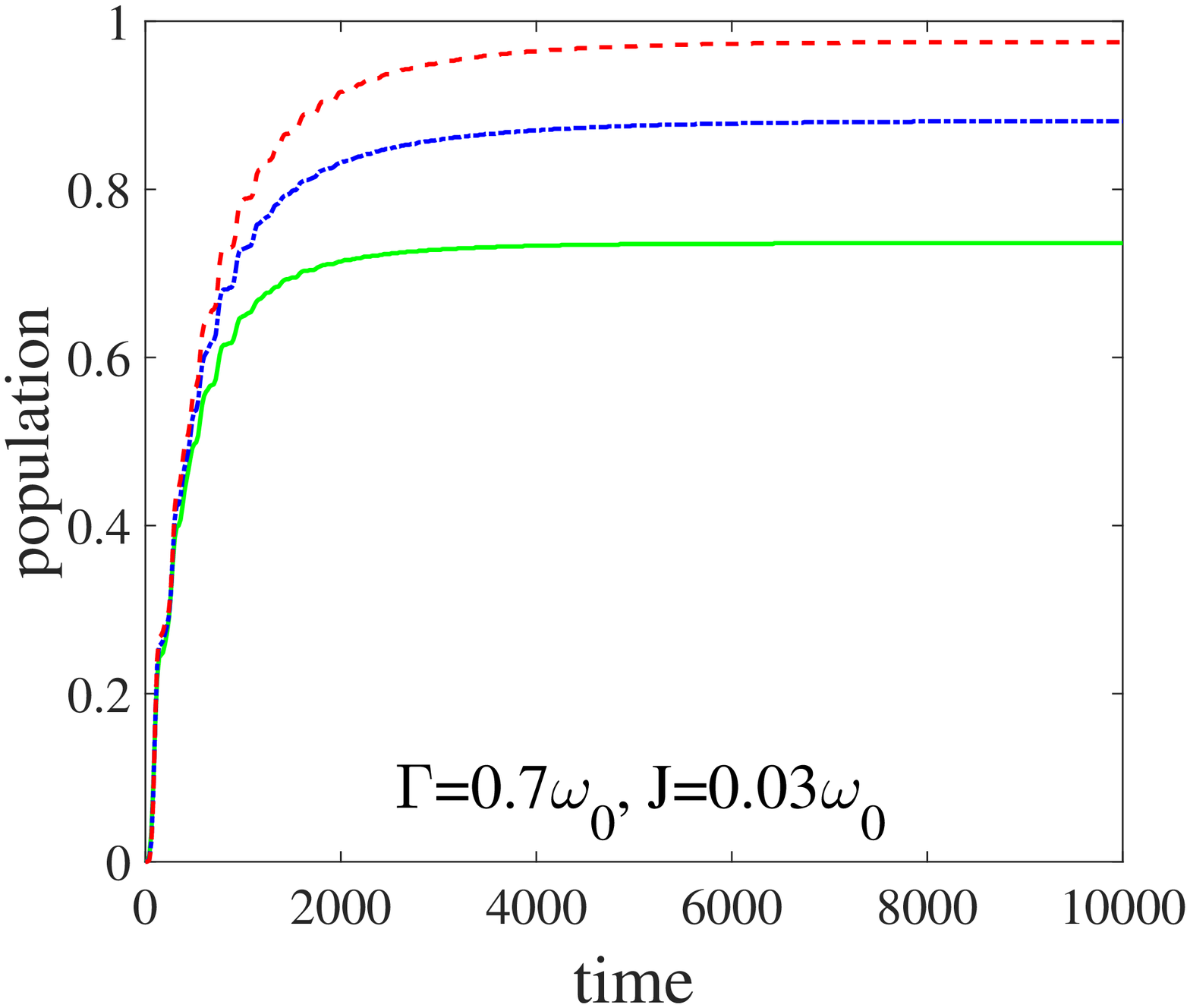}
        \label{fig:second_sub}
    }
    \caption{}
    \end{figure}

\end{document}